\documentclass{elsartModified1}  

\usepackage{graphicx}
\usepackage{epsfig}

\usepackage{latexsym,amssymb,amsfonts}  

\begin{document}

\begin{frontmatter}

\title{
Long time scale in Hamiltonian systems with internal degrees of freedom: 
Numerical study of a diatomic gas
}

\author{Yoshihiro Watanabe}\footnote{Email: yoshi@ccl.scc.kyushu-u.ac.jp}
\address{Department of Chemistry, Faculty of Sciences, Kyushu University\\
Hakozaki, Fukuoka 812-8581, Japan }

\author{Nobuko Fuchikami}\footnote{Corresponding author. 
Email: fuchi@phys.metro-u.ac.jp}
\address{Department of Physics, Tokyo Metropolitan University\\
1-1, Minami-Ohsawa, Hachioji, Tokyo 192-0397, Japan}

\begin{abstract}
We performed molecular dynamics simulations on a one-dimensional 
diatomic gas to 
investigate the possible long time scale inherent in heterogeneous 
Hamiltonian systems. 
The exponentially long time scale for energy sharing between 
the translational motion and the vibrational one certainly 
exists in a large limit of the system size.
The time scale depends on the vibrational frequency $\omega$ 
not as in the pure exponential form $\sim \exp[B\omega]$ 
but as $\sim \exp[B\omega^{\alpha}]$ with 
$\alpha<1$, in good agreement with the expression derived 
from the Landau-Teller approximation. 
The numerical simulations show that the complete resonance condition for 
vibrational frequencies assumed in the analytical treatment is not essential 
for the long time scale. 
Some discussions of $1/f$ fluctuations based on the present results 
will be given.

\end{abstract}

\end{frontmatter}

\section{Introduction}
\label{sec:level1}
The experimental fact that the high frequency degrees of freedom 
do not contribute to 
the heat capacity of molecular gases is a seeming 
violation of the equipartition theorem in the 
classical statistical mechanics. 
As a possible explanation of this 
effective ``freezing'' phenomenon, 
Boltzmann\cite{boltzmann1895} and Jeans \cite{jeans1903} suggested 
more than one century ago 
that, if the time scale 
in vibration and the time scale associated with a typical binary 
collision in the gas are largely different,
the amount of the energy exchanged 
between the vibrational and translational degrees of freedom can be 
small enough so that it takes an extremely long time for 
realization of thermal equilibrium.
Although this problem 
was solved by the arrival of quantum mechanics, 
considerable attention has been paid to their original idea 
(the Boltzmann-Jeans conjecture) as a possible scenario of the slow dynamics 
in heterogeneous systems since their viewpoint was first noticed
by Benettin et al. \cite{benettin87,benettin87(I),benettin89(II)}. 
For example, a modified Fermi-Pasta-Ulam model which is composed of 
two subsystems with different time scales (corresponding to
acoustic and optical vibrational modes) 
was analyzed to derive nonequipartition 
of energy in macroscopic systems \cite{galgani92}. 
$1/f$-type energy fluctuations observed in 
molecular dynamics (MD) simulations of 
liquid water \cite{sasai92} and the 
long-term energy storage in protein molecules \cite{ishijima98}
were also discussed 
in the light of 
the Boltzmann-Jeans conjecture \cite{shudo05,nakagawa01}. 

Generic nearly-integrable Hamiltonian systems 
have the stability time of the form 
\begin{equation}
\tau_{\, \rm N} \sim \exp[1/\varepsilon^{\alpha}] 
\qquad \mbox{for} \quad 0< \varepsilon < \varepsilon_0 
\quad \mbox{(the Nekhoroshev theorem),}
\label{tau_N}
\end{equation}
where $\varepsilon$ is a positive parameter characterizing the smallness of 
the perturbation joined to an unperturbed integrable Hamiltonian; 
$\alpha$ and $\varepsilon_0$ are positive constants depending on 
the explicit form of the Hamiltonian, especially on the system size, i.e. 
the number $N$ of degrees of freedom \cite{nekhoroshev77}. 
The time scale $\tau_{\, \rm N}$ 
grows exponentially as the perturbation parameter $\varepsilon$ becomes small.

Equation (\ref{tau_N}) explains the non-ergodic behavior discovered
in weakly coupled harmonic oscillators  by 
Fermi, Pasta and Ulam (FPU phenomenon) 
for finite $N$ \cite{FPU55}. 
However, 
the general proof based on the perturbational techniques yields 
a strong $N$-dependence for the constants: $\alpha \sim 1/N$ and 
$\varepsilon_0 \sim N^{-N}$ in the limit of $N \to \infty$ \cite{benettin86}, 
which implies that the FPU phenomenon will disappear in large systems or 
in the thermodynamic limit, although, 
as pointed out in \cite{benettin86}, 
this general results do not exclude the possibility 
that some specified system of the FPU-type could 
actually exhibit the non-ergodic behavior 
in the large $N$ limit \cite{footnote1}.

Another type of Hamiltonian systems which can exhibit the long time scale of 
the Nekhoroshev-type is presented in refs. 
\cite{benettin87(I),benettin89(II)}, 
the idea being based on the Boltzmann-Jeans conjecture. 
Such systems are represented by
\begin{equation}
H=H_0(p,q)+ H_1(\pi, \xi) + f(p,q,\pi, \xi), 
\label{H_Ben}
\end{equation}
where $H_0$ 
with the canonical variables 
$p \equiv (p_1, \cdots ,  p_{\nu})$ and $q \equiv (q_1, \cdots , q_{\nu})$ 
may generally be non-integrable in contrast to the unperturbed 
Hamiltonian in the Nekhoroshev theorem, 
while $H_1$ describes the system composed of $N$ harmonic 
oscillators: 
\begin{equation} 
H_1= \sum_{i=1}^{N} \left( \frac{\pi_i^2}{2}+
\frac{\omega_i^2\xi_i^2}{2}\right)
\label{H-1}
\end{equation}
with $\pi \equiv (\pi_1, \cdots , \pi_N)$ and 
$\xi \equiv (\xi_1, \cdots , \xi_N)$.
The interaction $f$ is assumed to vanish for $\xi=0$. 
The stability time is obtained as  
\begin{equation}
\tau_{\, \rm B} \sim \exp[\, (\lambda/\lambda^*)^{\alpha}\,],
\label{tau_B}
\end{equation}
where $\lambda \equiv \omega/\Omega$ is the ratio of two typical time scales, 
$1/\Omega$ in $H_0$ and $1/\omega$ in $H_1$ and $\lambda \gg 1$ is assumed. 
The positive constants $\alpha$ and $\lambda^*$ generally depend on 
the number $N$ of harmonic oscillators. 

When the angular frequencies $(\omega_1, \cdots , \omega_N)$ are nonresonant, 
the exponent $\alpha$ is
estimated again as $\alpha \sim 1/N$, but when 
$\omega_1=\cdots =\omega_{N} \equiv \omega$ (the complete resonance) holds, 
$\alpha=1$ is proved for any $N$, so that the long time scale of 
the pure exponential form
\begin{equation}
\tau_{\, \rm B} \sim e^{\lambda/\lambda^*} \equiv e^{B\omega}
\label{tau-B-J}
\end{equation}
is expected even for macroscopic systems as 
originally suggested by Jeans. 
If the stability time of the above form remains finite for $N \to \infty$, 
the basic idea of Boltzmann-Jeans could work as a key 
to reveal the dynamics of slow relaxation observed in macroscopic systems. 
The situation is, however, not so simple 
because the common perturbational techniques yield strong 
$N$-dependence of $\lambda^*$ as $\lambda^* \sim N^2$ 
even in the complete resonance case \cite{benettin94} 
and the long time scale cannot be guaranteed in the thermodynamic limit. 
Here too this does not 
mean 
the long time scale will always disappear for large systems. 
In fact, the above estimate of $N$-dependence 
seems to be
statistically unacceptable 
if applied to sufficiently dilute molecular gases 
in which only binary collisions are relevant. 
Unfortunately, it is not easy to draw 
definite conclusions about the stability time 
(whether it remains finite or tends to zero in the thermodynamic limit) 
and its more reliable $\omega$-dependence (whether it is pure exponential 
or else) from direct simulations based on the molecular dynamics 
because the computational time steeply increases with the system size and 
the vibrational frequency, though 
such simulations were attempted for a diatomic gas
nearly two decades ago \cite{benettin87}. 

Instead of invoking direct simulations, the $\omega$-dependence 
is considered from a different angle in \cite{benettin99}: 
By assuming a gas composed of identical diatomic molecules (with 
$\omega_1= \cdots  =\omega_N = \omega$)
and by applying the Landau-Teller approximation \cite{landau36,rapp60} to 
binary collisions between molecules 
the $\omega$-dependence of the time scale is obtained as 
\begin{equation}
\sim \exp[B \omega^{\alpha}]  \quad \mbox{with} \quad \alpha\equiv 
\frac{2}{3+2/s} < \frac{2}{3}, 
\label{tau-stretch-exp}
\end{equation}
where the inter-molecular potential $\phi(r)$ 
is of the form 
\begin{equation}
\phi(r) \sim \frac{1}{r^s}, \quad s\ge 1 \quad \mbox{for small $r$}.
\label{phi-r}
\end{equation}

In the present paper, we perform microcanonical MD simulations 
of a one-dimensional diatonic gas as a minimal model of
macroscopic heterogeneous systems (i.e. systems with internal degrees of 
freedom) and investigate its long-term behavior 
to supplement with some information which has not been presented so far, 
hoping to find a clue to basic understanding of 
the slow relaxation phenomena observed in more realistic systems. 

We first confirm whether the long time scale remains or not in the 
large $N$ limit, which we think is necessary because the time scale 
does decrease with increasing $N$ in a range of small $N$ (for a fixed value 
of the number density). 
Next, we 
obtain its $\omega$-dependence to examine 
the Landau-Teller approximation. 
Since both the perturbation method leading to Eq. (\ref{tau-B-J}) and the 
Landau-Teller approximation applied to the binary collisions leading to 
Eq. (\ref{tau-stretch-exp}) assume 
the condition of the complete resonance, 
we finally test the case in which 
$(\omega_1, \cdots , \omega_N)$ are slightly different with each other.

\section{Hamiltonian and Initial Condition}

We consider a one-dimensional molecular gas composed of $N$ identical 
diatomic molecules with mass $M$ \cite{footnote2}.
The positions of two atoms belonging to the $i$th molecule 
are denoted by 
$q_i-a-\xi_i$ and $q_i+a+\xi_i$, where $q_i$ is 
the center of mass of the molecule 
and $2a$ is the bond length, $\pm \xi_i$ are 
the displacements of atoms from the equilibrium position.
The total Hamiltonian is given by 
\begin{equation}
H =N \left[K_{\rm tr}(p)+ K_{\rm vib}(\pi)+V_{\rm vib}(\xi)
+V_{\rm int}(q, \xi) \right], 
\label{H-diatom}
\end{equation}
where
\begin{equation}
K_{\rm tr}(p) = \frac{1}{N}\sum_{i=1}^{N}\frac{p_i^2}{2M} \quad 
\mbox{with} \quad 
p_i\equiv M \frac{dq_i}{dt}
\label{K-tr}
\end{equation}
is the kinetic energy 
of translational motion of molecules,
\begin{equation}
K_{\rm vib}(\pi) =
 \frac{1}{N} \sum_{i=1}^{N}\frac{\pi_i^2}{2M} \quad \mbox{with} \quad 
\pi_i\equiv M \frac{d \xi_i}{dt}
\label{K-vib}
\end{equation}
and
\begin{equation}
V_{\rm vib}(\xi) =
\frac{1}{N} \sum_{i=1}^{N}\frac{M \omega^2 \xi_i^2}{2}
\label{V-vib}
\end{equation}
are the kinetic and potential energies of intra-molecular vibrations, 
respectively, and
\begin{equation}
V_{\rm int}(q, \xi)=
\frac{1}{N} \sum_{i=1}^{N} 
\phi \left( (q_i-a-\xi_i)-(q_{i-1}+a+\xi_{i-1}) \right)
\label{V-int}
\end{equation}
is the inter-molecular potential. 
We employ the cyclic boundary condition: 
$q_{0}\equiv q_{N}$ and 
$\xi_{0}\equiv \xi_{N}$.
The molecular interaction is identical with that adopted in
\cite{benettin87}, which is short range repulsive and given by  
\begin{equation}
\phi(r) = \phi_0 \frac{e^{-(r/r_0)^2}}{r/r_0}, \qquad r>0.
\label{pot-phi-r}
\end{equation}

In the present system, 
$N\left[K_{\rm tr}(p)+V_{\rm int}(q,0)\right]$ corresponds to $H_0(p,q)$ 
with $\nu=N$ and  
$N\left[V_{\rm int}(q,\xi)\right.$ $\left.-V_{\rm int}(q,0)\right]$ to 
$f(p,q,\pi, \xi)$ in Eq. (\ref{H_Ben}). 
We may set 
$M=r_0=\phi_0=1$ without losing generality by choosing 
the units of mass, length and energy as $M$, $r_0$ and $\phi_0$, respectively. 
Then the unit of time is $ r_0 \sqrt{M/\phi_0} \equiv 2\pi/\Omega$, 
which is a 
typical time scale characterizing the Hamiltonian $H_0(p,q)$. 
Note that $\omega$ scaled by this time scale is 
actually read as the time scale $ratio$ $\omega/\Omega$ of the two subsystems. 

Not like many 
FPU-type simulations starting from a state in which the energy is 
distributed to only specific normal modes, 
we chose a nearly thermal equilibrium state as the 
initial state, except in Section \ref{subsec:from-noneq}. 
This is because 
we are mainly interested in the long-term behavior observed in 
macroscopic systems in thermal equilibrium.

If the system represented by the Hamiltonian (\ref{H-diatom}) is 
in a thermal equilibrium state with the temperature $T$, 
the expectation value of the kinetic energy $K_{\rm tr}$ 
(in the canonical ensemble) is calculated as
\begin{eqnarray}
\left<K_{\rm tr}(p) \right> &=& \frac{\frac{1}{N}
\sum_{i=1}^N \int \frac{p_i^2}{2M} e^{-H/kT} dq d\xi dp d\pi}
{\int e^{-H/kT} dq d\xi dp d\pi} \nonumber \\
 &=& 
 \frac{1}{N}
\sum_{i=1}^N \frac{\int \frac{p_i^2}{2M} e^{-p_i^2/2M kT} dp_i}
{\int e^{-p_i^2/2M kT} dp_i} = \frac{kT}{2}.
\label{kitaichi-K-tr}
\end{eqnarray}
Note that Eq. (\ref{kitaichi-K-tr}) is 
exact even if the interaction term $V_{\rm int}(q, \xi)$ exists, 
because the variable $p_i$ is separated from the rest of the
variables. 
Similarly, one obtains
\begin{equation}
\left<K_{\rm vib}(\pi) \right> = \frac{kT}{2}, 
\label{kitaichi-K-vib}
\end{equation}
which is also exact.

Without the term $V_{\rm int}$, the Hamiltonian is decoulped into three 
terms with $N$ degrees of 
freedom for each:
$NK_{\rm tr}(p), NK_{\rm vib}(\pi)$, and $NV_{\rm vib}(\xi)$; 
and the expectation value of $V_{\rm vib}(\xi)$ in the 
thermal equilibrium state 
is also obtained rigorously as
\begin{equation}
\left<V_{\rm vib}(\xi) \right> = 
\frac{kT}{2},
\end{equation}
namely	, the energy is equally distributed among three 
``subsystems'' as
\begin{equation}
\left< N K_{\rm tr}(p) \right> =
\left< N K_{\rm vib}(\pi) \right> = 
\left< N V_{\rm vib}(\xi) \right> = 
\frac{\left<H\right>}{3}.
\label{H/3}
\end{equation}

The present system is isolated (not contacted 
with the heat bath) and includes the coupling term 
$V_{\rm int}$. 
However, 
if the coupling energy is relatively small and $N$ is large, 
the energy of each subsystem in the stationary state 
is expected to be approximately equal, namely 
the long-time average for each subsystem satisfies 
\begin{equation}
N \overline{K_{\rm tr}(p)} \sim N \overline{K_{\rm vib}(\pi)} 
\sim N \overline{V_{\rm vib}(\xi)} 
\sim \frac{N E_{\rm tot}}{3},
\label{E/3}
\end{equation} 
where $E_{\rm tot}$ is the specific energy, i.e. the total energy of 
the system per molecule. 

Taking account of the above, we start simulations with an initial state 
which is 
$approximately$ thermal equilibrium as follows:

For the coordinate $\xi$, 
we generate 
$N$ independent random numbers $R_{\xi,i}$ $(i=1,2,\cdots, N)$
with the normal distribution 
(Gaussian distribution with zero mean and unit variance)
 and set the initial value as $\xi_i(0)=S_{\xi} R_{\xi,i}$, where 
the positive constant $S_{\xi}$ is chosen so that
\begin{equation}
V_{{\rm vib},0}\equiv \frac{1}{N}
\sum_{i=1}^N \frac{M\omega^2 \xi_i(0)^2}{2} 
= \frac{E_{\rm tot}}{3}
\label{init-V-vib}
\end{equation}
holds. This condition 
corresponds to 
Eq. (\ref{E/3}), i.e. a thermal equilibrium state.

The initial position of the center of mass for each molecule is chosen as 
\begin{equation}
q_i(0)=(i-1)L+S_q R_{q,i}, 
\label{init-q-i}
\end{equation}
where $L$ is the average molecular distance, $S_q R_{q,i}$ is a small 
deviation from the average and 
$R_{q,i}$ $(i=1, \cdots, N)$ are   
$N$ independent random numbers with the normal distribution.
The constant $S_q$ is a small positive 
number so that the condition 
$q_i(0) < q_{i+1}(0)$ ($i=1, \cdots, N$) is satisfied. 
The initial value of the molecular interaction energy
\begin{equation}
V_{{\rm int},0}\equiv V_{\rm int}\left(q(0), \xi(0)\right)
\label{init-V-int}
\end{equation}
is expected to be small by choosing the inter-molecular distance 
$L$ being large enough in comparison to the potential 
parameter $r_0$.

Next, we choose the initial momentum $p_i(0)$ ($i=1, \cdots, N$) 
by a similar way as for $\xi_i(0)$ but with a constraint 
$\sum_{i=1}^{N}p_i(0)=0$: 
We generate $N$ independent random numbers $R_{p,i}$ with the normal 
distribution \cite{footnote3}
and set 
$p_i(0)= S_p \left(R_{p,i}-\sum_{j=1}^{N} R_{p,j}/N \right)$ with 
a constant $S_p$ which satisfies
\begin{equation}
K_{{\rm tr},0} \equiv
\frac{1}{N}\sum_{i=1}^N \frac{p_i(0)^2}{2M} 
= \frac{E_{\rm tot}-V_{{\rm vib},0}-V_{{\rm int},0}}{2}.
\label{init-K-tr}
\end{equation}
Then $K_{{\rm tr},0}$ is almost equal to $E_{\rm tot}/3$ 
when $V_{{\rm int},0}$ 
is small.

Finally $\pi_{i}(0)$ ($i=1,2,\cdots, N$) are chosen from              
$N$ independent random numbers $R_{\pi ,i}$ with normal distribution and 
set $\pi_i(0)=S_{\pi} R_{\pi,i}$ with a constant $S_{\pi}$ satisfying 
\begin{equation}
K_{{\rm vib},0} \equiv 
\frac{1}{N}\sum_{i=1}^N \frac{\pi_i(0)^2}{2M} = 
K_{{\rm tr},0}.
\label{init-K-vib}
\end{equation} 
Hereafter we mean the initial state satisfying 
Eqs. (\ref{init-V-vib}) $\sim$ (\ref{init-K-vib}) by 
``nearly equilibrium initial state''. 
Since the parameters $L$ and $a$ appear only
in the form $L-2a$ in $V_{\rm int}$ (see Eq. (\ref{V-int})), we may set $a=0$ 
by including its effect in the parameter $L$.

\section{Numerical Results}
The computer simulations were performed using the symplectic 
method \cite{yoshida90}. As mentioned in the previous section, 
We set $M=r_0=\phi_0=1$. 
Throughout the present simulations the average molecular distance 
was fixed as $L=4$. 
\subsection{Time evolution}
\begin{figure}
\scalebox{0.55}{\includegraphics{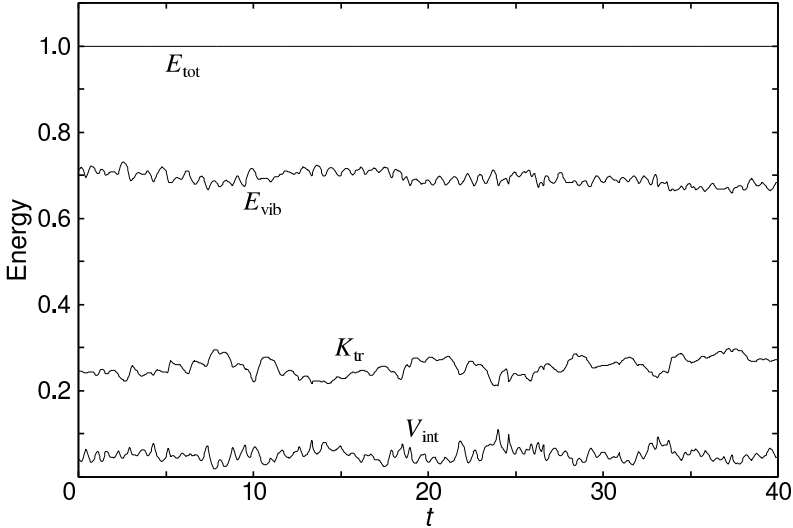}}
\caption{\label{ShortPlotE}
A sample of the temporal fluctuations of energies for 
$N=128$ and $\omega=8$. 
$E_{\rm tot}$ :
the total energy (per molecule), 
$K_{\rm tr}$：   
the kinetic energy of the translational motion,
$E_{\rm vib}=K_{\rm vib}+V_{\rm vib}$：
the vibrational energy, 
$V_{\rm int}$：the molecular interaction energy．
The initial condition: 
$E_{{\rm tot},0}=1, V_{{\rm vib},0}=1/3, V_{{\rm int},0}=0.000\; 024, 
K_{{\rm tr},0}=K_{{\rm vib},0}=0.333\; 322$. 
Instantaneous values are plotted with 
the sampling time interval $\Delta t=0.04$ 
in the interval $[230, 230+40]$ 
after the system was relaxed for $t_{\rm relax}=230$. 
}
\end{figure}

Employing the initial condition in Eqs. (\ref{init-V-vib}) 
$\sim$ (\ref{init-K-vib}), we started 
the simulation from a nearly thermal equilibrium state, in which 
$K_{\rm tr}(p)= K_{\rm vib}(\pi)\approx V_{\rm vib}(\xi)= E_{\rm tot}/3$ 
and $V_{\rm int} \approx 0$.
The time dependence of energies of the total system (per molecule) 
for a short time interval 
$230 \le t \le 230+40$ 
is presented in Fig. \ref{ShortPlotE} for $N=128$ and $\omega=8$, in which 
\begin{equation}
E_{\rm tot}\equiv K_{\rm tr}+E_{\rm vib}+V_{\rm int}
\label{Etot}
\end{equation}
and 
\begin{equation}
E_{\rm vib}\equiv K_{\rm vib}+V_{\rm vib}.
\label{Evib}
\end{equation}

\begin{figure}
\scalebox{0.55}{\includegraphics{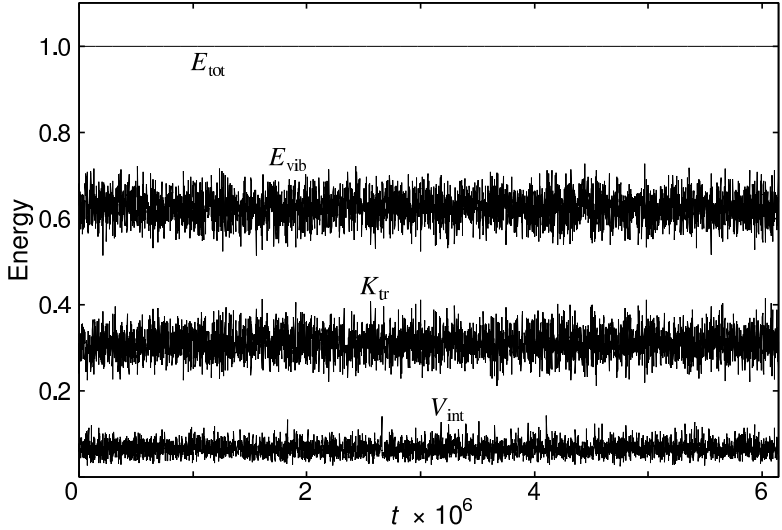}}
\caption{\label{LongPlotE}
Long time behavior. 
The same system parameters and the same initial condition as in 
Fig. \ref{ShortPlotE}.
Instantaneous values are plotted 
at 4096 time points with the sampling interval $\Delta t=1500$.
}
\end{figure}

\begin{figure}
\scalebox{0.55}{\includegraphics{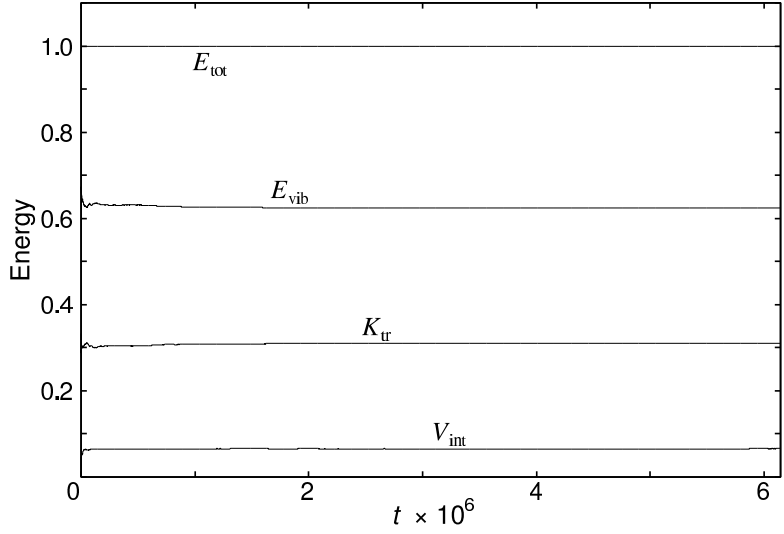}}
\caption{\label{AveE}
Long time behavior of energy fluctuations. 
The same run as in Fig. \ref{LongPlotE}, but 
the plotted point at the time $t$ represents the average over the 
time interval $[0, t]$ defined by (\ref{time-ave}). 
The sampling interval is the same as in Fig. \ref{LongPlotE}.
}
\end{figure}

\begin{figure}
\scalebox{0.55}{\includegraphics{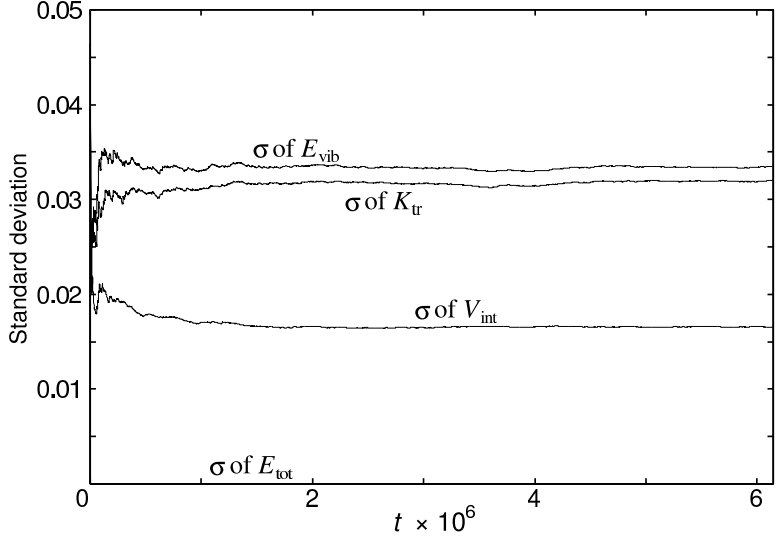}}
\caption{\label{SigmaE}
Standard deviation $\sigma$ of energies corresponding to 
Fig. \ref{AveE}. 
Plotted points at the time $t$ are obtained from (\ref{Sigma-t}). 
The sampling interval is the same as in Fig. \ref{LongPlotE}.
The standard deviation of the total energy,  
$\sigma$ of $E_{\rm tot}$, is $1.2 \times 10^{-5}$. 
}
\end{figure}

Figures \ref{LongPlotE}, \ref{AveE} and \ref{SigmaE} show
the long time behavior of energies and their standard deviations. 
In Fig. \ref{AveE}, 
the plotted point at the time $t$ is the average of the 
quantity $A$ defined by  
\begin{equation}
\overline{A(t)} \equiv 
\frac{1}{t}\int_{0}^{t} A(t')dt'.
\label{time-ave}
\end{equation}
In Fig. \ref{SigmaE}, the standard 
deviation at the time $t$ for the quantity $A$ is plotted, which is defined by
\begin{equation}
\sigma(t) \equiv 
\sqrt{\overline{A(t)^2}-\left(\overline{A(t)}\right)^2}. 
\label{Sigma-t}
\end{equation}

Each energy and its standard deviation are plotted with 
the sampling time interval $\Delta t=1500$.
The final values of energies in Fig. \ref{AveE} are 
$E_{\rm tot}=1.000\; 001 \; 63$, 
$K_{\rm tr}=0.310\; 09$, 
$E_{\rm vib}=0.312\;39 + 0.311\;86=0.624\;25$,  
$V_{\rm int}=0.065\;66$. 
Those values agree well with 
the thermal equilibrium energies
$K_{\rm tr}=(E_{\rm tot}-V_{\rm int})(N-1)/(3N-1)=0.309\;82$, 
$K_{\rm vib}=V_{\rm vib}=(E_{\rm tot}-V_{\rm int})N/(3N-1)=0.312\;26$ for 
$N=128$. 
The factors $(N-1)/(3N-1)$ and $N/(3N-1)$ instead of $1/3$ appear because 
the center of mass of the total system is at rest. 
In Fig. \ref{SigmaE}, the final value of 
the standard deviation $\sigma$ of the total energy $E_{\rm tot}$ 
is $1.2 \times 10^{-5}$, 
which is a measure of the computational error. 

\begin{figure}
\scalebox{0.55}{\includegraphics{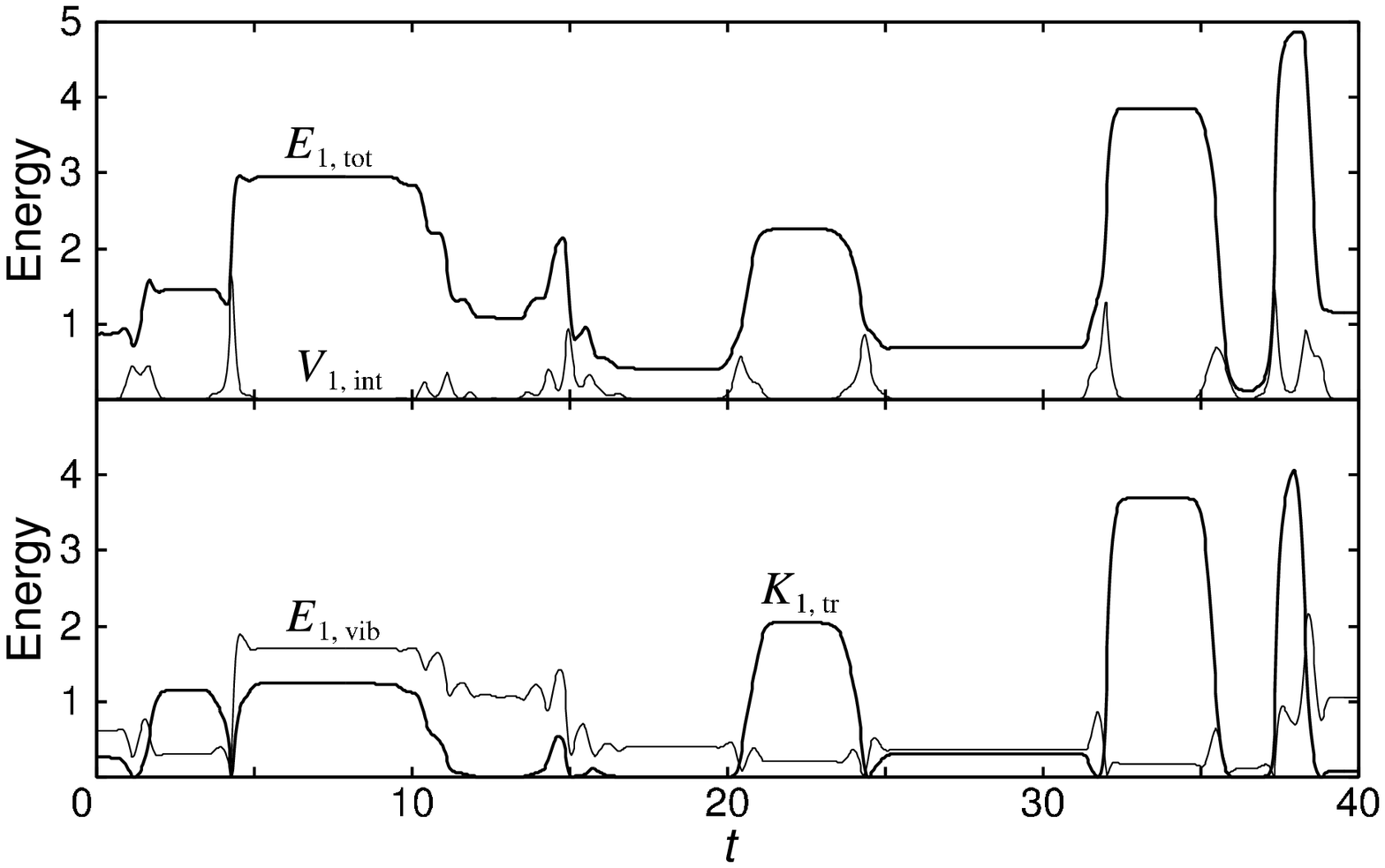}}
\caption{\label{MoleculeE}
Energies of molecule 1 in the 
same run as in Fig. \ref{ShortPlotE}.  
$K_{1, {\rm tr}}=p_1^2/2M$, 
$E_{1, {\rm vib}}=\pi_1^2/2M+M\omega^2\xi_1^2/2$, 
$V_{1, {\rm int}}=(\phi(q_1-q_0-\xi_1-\xi_0)+\phi(q_2-q_1-\xi_2-\xi_1))/2$, 
$E_{1, {\rm tot}}=K_{1, {\rm tr}}+E_{1, {\rm vib}}+V_{1, {\rm int}}$. 
}
\end{figure}

\begin{figure}
\scalebox{0.55}{\includegraphics{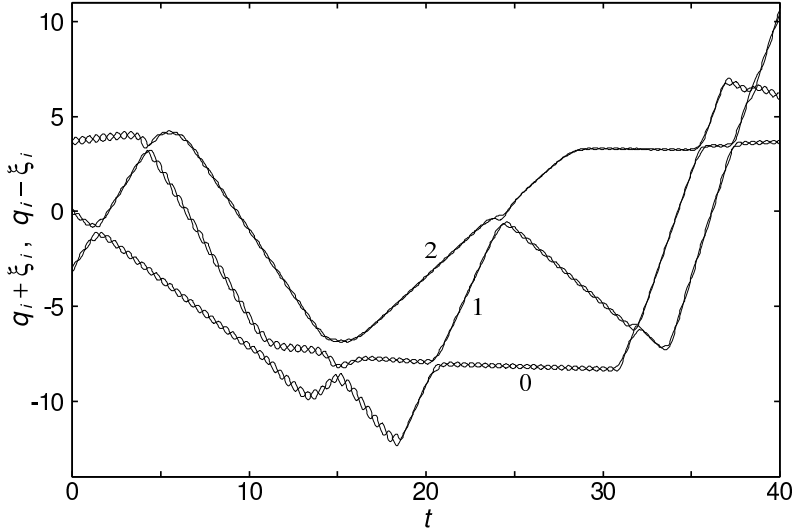}}	
\caption{\label{q1Plot}
Motion of atoms belonging to 
molecules 0, 1 and 2. 
The coordinates of atoms $q_i\pm \xi_i \, (i=0,1,2)$ are plotted in the 
same run as in Fig. \ref{ShortPlotE}.  
}
\end{figure}

It should be noted that 
the energies of individual molecules change quite largely 
(compare with $E_{\rm tot}=1$ which is averagely distributed to 
each molecule) as can be seen in 
Fig. \ref{MoleculeE} in which instantaneous 
energies of molecule 1 are plotted. 
The large energy exchange is induced by the collision with the 
neighboring molecules as seen from Fig. \ref{q1Plot}.

\subsection{Size-dependence of the correlation time}

As a time scale characterizing the system in equilibrium, 
one may employ the correlation time of the energy fluctuation. 
Following \cite{benettin87}, we computed the correlation time 
$\tau_{\rm cor}$ of the vibrational energy $E_{\rm vib}$
which is defined such that the normalized correlation 
function 
\begin{equation}
G(\tau) \equiv \frac{\overline {E_{\rm vib}(t)E_{\rm vib}(t+\tau)}-
\left(\overline {E_{\rm vib}(t)}\right)^2}{\overline {E_{\rm vib}(t)^2}-
\left(\overline {E_{\rm vib}(t)}\right)^2}
\end{equation} 
becomes a half of the initial value $G(0)=1$: 
\begin{equation}
G(\tau_{\rm cor})=0.5,
\label{def-tau}
\end{equation}
where the overline denotes  
the time average over the dummy variable $t$ for a sufficiently large 
interval $t_{\rm obs}$.
 
We computed $\tau_{\rm cor}$ for various samples with different random 
initial conditions satisfying 
Eqs. (\ref{init-V-vib}) $\sim$ 
(\ref{init-K-vib})  
and define the average by
\begin{equation}
\overline {\tau}_{\rm cor} \equiv {\rm Ave}\left[\tau_{\rm cor}\right].
\label{tau-bar}
\end{equation} 

\begin{figure}
\scalebox{0.55}{\includegraphics{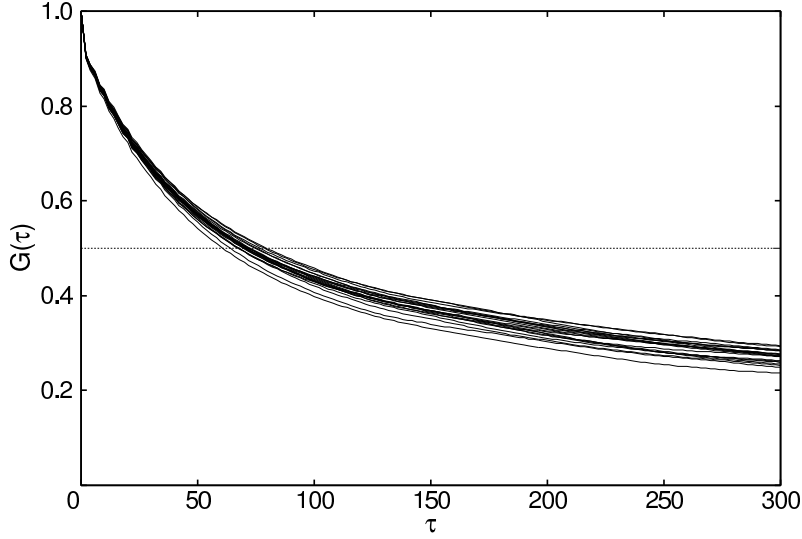}} 
\caption{\label{G-tau}
Correlation function $G(\tau)$ of the vibrational energy 
$E_{\rm vib}$ for 
20 independent runs with nearly equilibrium initial states. 
$N=128$, $\omega=8$, $E_{\rm tot}=1$.
}
\end{figure}
We have also employed another averaged correlation time $\tau_{\rm cor}'$ 
which is defined such that 
\begin{equation}
\overline{G}(\tau_{\rm cor}') \equiv 0.5,
\label{tau-prime}
\end{equation} 
where $\overline{G}(\tau)$ is 
the correlation function averaged over the samples:
\begin{equation}
\overline{G}(\tau) \equiv {\rm Ave} \left[G(\tau)\right].
\end{equation}
The correlation time $\tau_{\rm cor}'$ is more easily estimated especially when $G(\tau)$ decays very slowly in some samples and it takes very long time to obtain the average $\overline{\tau}_{\rm cor}$ over all samples, 
which actually occurs for small $N$.

The correlation function $G(\tau)$ is plotted in Fig. \ref{G-tau} 
for $N=128$, $\omega=8$, $E_{\rm tot}=1$ and for 20 samples with different 
random initial states 
satisfying (\ref{init-V-vib}) $\sim$ (\ref{init-K-vib}). 
To estimate $G(\tau)$ reliably over the range $0 \le \tau \le 300$, 
the time average was taken in the 
interval $[t_{\rm relax}, t_{\rm relax} + t_{\rm obs}]$ 
with the observation time 
$t_{\rm obs} =1 \times 10^{6}$ after the system 
was relaxed for $t_{\rm relax} = 2\times 10^{4} $. 
\begin{figure}
\scalebox{0.55}{\includegraphics{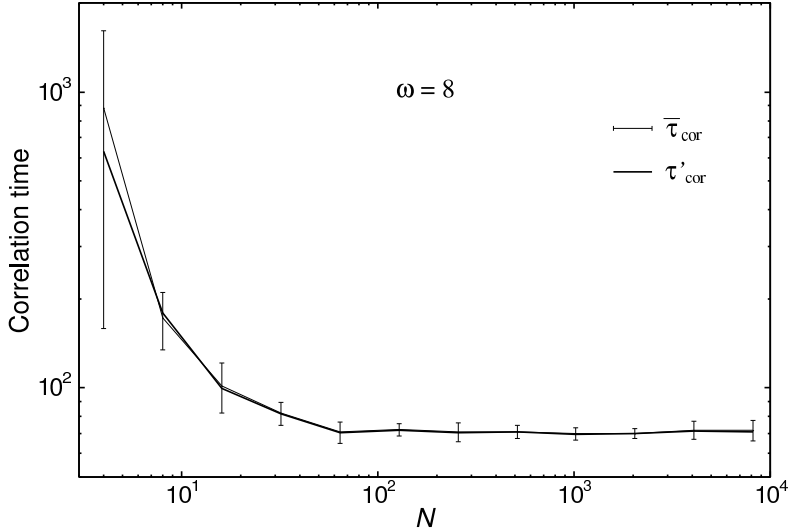}} 
\caption{\label{tau-N}
System size dependence of the time scale. 
Two kinds of correlation time $\overline{\tau}_{\rm cor}$ defined by 
(\ref{tau-bar}) and $\tau_{\rm cor}'$ by eq. (\ref{tau-prime}) are 
plotted against the number $N$ of molecules. 
The average is taken over 20 samples, i.e. 20 independent runs 
starting from nearly equilibrium initial states.
The error bars indicate $\overline{\tau}_{\rm cor} \pm \sigma$, 
where $\sigma$ is the standard deviation for 20 samples. 
$\omega=8$, $E_{\rm tot}=1.0$. 
}
\end{figure}

In Fig. \ref{tau-N}, $\overline{\tau}_{\rm cor}$ and 
$\tau_{\rm cor}'$ are plotted against the system size, i.e. 
the number $N$ of molecules, from $N=4$ to $N=8192$, where $\omega=8$ and 
the specific energy is fixed at $E_{\rm tot}=1.0$. 
The average is obtained from 20 independent runs 
starting from nearly equilibrium initial states 
and the error bars indicate $\pm \sigma$, $\sigma$ 
being the standard deviation of $\tau_{\rm cor}$ for 20 runs..
The correlation time as a function of $N$ decreases with increasing $N$ 
in the beginning but tends to a finite value $\approx 71$ as $N \to \infty$.  
The difference between $\overline{\tau}_{\rm cor}$ and $\tau_{\rm cor}'$ is 
large for small values of $N$ but very small for large $N$.
The error bar of $\overline{\tau}_{\rm cor}$ 
also 
becomes small for large $N$ if the precision of the numerical simulation 
is high enough. 
Figure \ref{tau-N} proves that the correlation time certainly stays finite 
for large $N$, where the molecular density is fixed at $1/L=1/4$. 

We have obtained a similar size dependence for a larger value of 
the specific energy: 
When $E_{\rm tot}=2.0$ with $\omega=8$, both 
$\overline{\tau}_{\rm cor}$ and $\tau_{\rm cor}'$ tend to 
$\approx 10.9$ with increasing $N$ as 
$(N, \overline{\tau}_{\rm cor} \pm \sigma, \tau_{\rm cor}')=$
$(8,13.6 \pm 4.7, 13.0),$ $(16, 11.0 \pm 1.0,10.9),$ 
$(32, 11.1 \pm 0.6, 11.0),$ $(64, 11.0 \pm 0.3, 11.0),$ 
$(128, 10.9 \pm 0.2,11.0),$ $(256, 10.9 \pm 0.1, 11.0),$ 
$(512, 10.9 \pm 0.1, 10.9)$.
For larger values of $E_{\rm tot}$, 
naturally the time scale is smaller 
but reaches a constant value faster.
Similar dependence of the time scale on the system size and on the 
specific energy 
has been reported from the numerical simulations of 
FPU $\beta$-model \cite{berchialla04}.

\subsection{$\omega$-dependence of the correlation time}

\begin{figure}
\scalebox{0.55}{\includegraphics{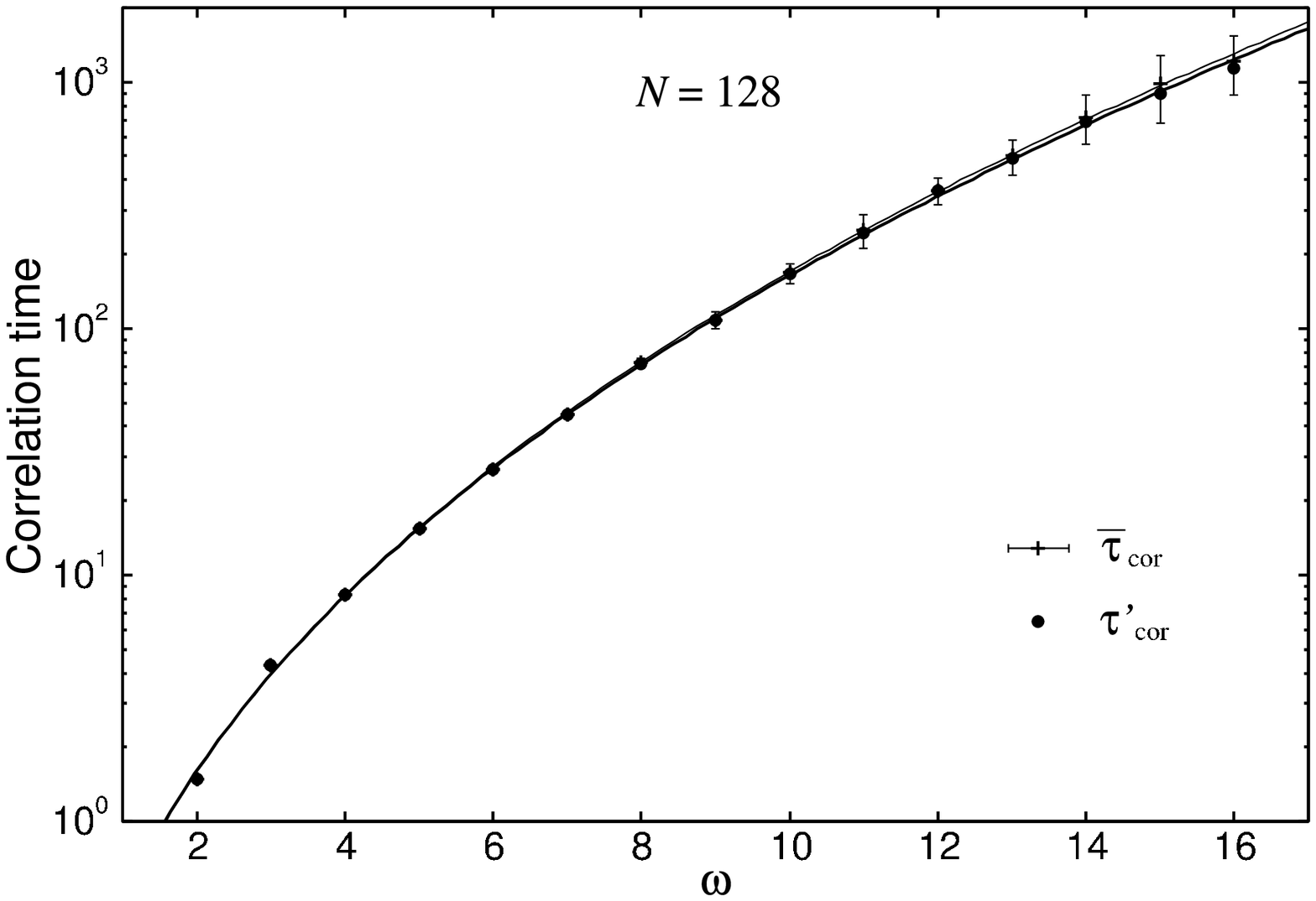}} Fig.9
\caption{\label{tau-omega-128}
Semilog plot of
$\omega$-dependence of the averaged correlation times
$\overline{\tau}_{\rm cor}$ and $\tau_{\rm cor}'$.
The average is taken over 20 independent runs.
Error bars indicate $\overline{\tau}_{\rm cor} \pm \sigma$, 
where $\sigma$ is the standard deviation for 20 runs 
with nearly equilibrium initial states.
The fitting lines are  
$\overline{\tau}_{\rm cor}=0.0090\times \exp[3.92 \times \omega^{\alpha}]$ 
and 
$\tau_{\rm cor}'=0.0096 \times \exp[3.88 \times \omega^{0.4}]$ 
with 
$\alpha=0.4$. 
$N=128, E_{\rm tot}=1.0$.
}
\end{figure}

\begin{figure}
\scalebox{0.55}{\includegraphics{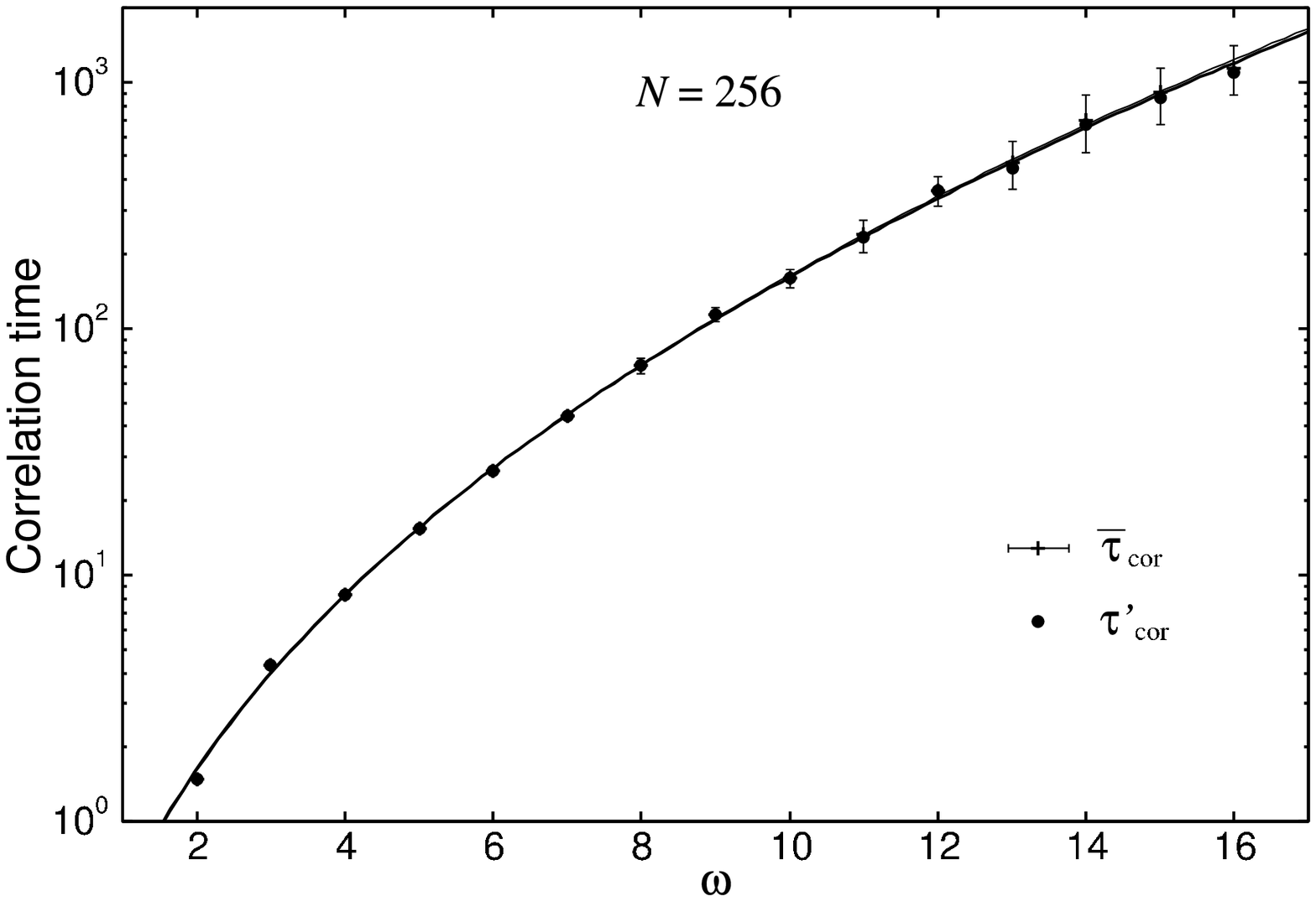}} Fig.10
\caption{\label{tau-omega-256}
The same as in Fig. {\ref{tau-omega-128}} but for $N=256$.
The fitting lines are  
$\overline{\tau}_{\rm cor}=0.0096 \times \exp[3.88 \times \omega^{\alpha}]$ 
and 
$\tau_{\rm cor}'=0.102 \times \exp[3.85 \times \omega^{\alpha}]$
with $\alpha=0.4$. }
\end{figure}

In Fig. \ref{tau-omega-128}, the averaged correlation times
$\overline{\tau}_{\rm cor}$ 
and $\tau_{\rm cor}'$ are 
plotted against $\omega$ for $N=128$. 
If the mathematical theorem (\ref{tau-B-J}) is applied directly 
to the present system in which the condition 
$\omega_1= \cdots =\omega_N=\omega$ is satisfied,  
the plot should drop on the straight line: 
$\sim \exp[B \omega^{\alpha}]$, with $\alpha=1$, 
while the present results can be fitted better with $\alpha$ smaller 
than one. 
The lines are for $\alpha=0.4$, which was suggested from 
Eq. (\ref{tau-stretch-exp}) (the Landau-Teller approximation) 
with $s=1$ corresponding to $\lim_{r\to 0}\phi(r) \sim 1/r^s$ for 
the potential $\phi(r)$ in Eq. (\ref{pot-phi-r}). 
We obtained the fitting curves 
$\overline{\tau}_{\rm cor}=A\; \exp[B  \omega^{0.4}]$ 
with
$(A,B)=(0.0090, 3.92)$ 
and $\tau_{\rm cor}'=A' \; \exp[B'  \omega^{0.4}]$ 
with $(A',B')=(0.0096, 3.88)$ \cite{footnote-tau-fit}. 
Also, Fig. \ref{tau-omega-256} shows good fitting to the same 
$\omega$-dependence in case of $N=256$ with
$(A,B)=(0.0096, 3.88)$ 
for $\overline{\tau}_{\rm cor}$ 
and $(A', B')=(0.0102, 3.85)$ for $\tau_{\rm cor}'$. 

Deviation of the $\omega$-dependence 
from the pure exponential form
can be qualitatively 
understood as follows \cite{benettin99}: 
The energy transfer $\Delta E$ from the translational motion to the vibrational one arises dominantly from direct binary collisions of 
molecules. 
The two subsystems of translational and vibrational degrees of freedom 
are assumed to be in equilibrium states with well-defined temperatures 
$T_{\rm tr}$ and $T_{\rm vib}$, respectively during 
the single binary collision. 
The Landau-Teller approximation applied to a single binary 
collision yields $\Delta E$ as 
a function of the asymptotic data of 
two molecules before the collision, which is 
of the form 
$\Delta E \sim e^{-\tau \omega}$. 
The parameter $\tau$ depends on 
the asymptotic value $p$ (at $t=-\infty$) of the relative momentum of 
the molecules, or equivalently on the 
translational 
energy $E_{\rm tr}\sim p^2/2$, 
and is expressed as $\tau \sim (E_{\rm tr})^{-\gamma}$ 
with $\gamma =(s+2)/2s>0$ for 
the inter-molecular potential (\ref{phi-r}).
The average energy exchange per unit time is given by 
integrating $\Delta E$ over various asymptotic states with   
the Boltzmann factor 
$\exp[-E_{\rm tr}/ k T_{\rm tr}-E_{\rm vib}/kT_{\rm vib}]$. 
Then 
the final form of the main $\omega$-dependence of the average energy exchange 
per unit time 
becomes stretched exponential: 
\begin{eqnarray}
\left<\Delta E\right> &\sim &
\int_0^{\infty} \exp \left[-(E_{\rm tr})^{-\gamma} \omega \right]
\exp \left[-E_{\rm tr} /k T_{\rm tr}\right] \, dE_{\rm tr}\\ \nonumber
&\sim & \exp[-B \omega^{\alpha}] 
\end{eqnarray}
with
\begin{equation}
\alpha=1-\frac{\gamma}{1+\gamma}=\frac{2}{3+2/s},
\end{equation}
which leads to the time scale $\sim 1/\left<\Delta E\right>$ as in 
Eq. (\ref{tau-stretch-exp}).
The origin of the exponent $\alpha$ smaller than one 
arises from the balance of the 
negative exponent $-\gamma$ in the parameter $\tau$, (i.e. 
the energy exchange $\Delta E \sim e^{-\tau \omega}$ in 
individual collisions increasing with $E_{\rm tr}$) 
and the weight of the Boltzmann factor decreasing with $E_{\rm tr}$.

\subsection{Power spectra and various time scales}
\begin{figure}
\scalebox{0.55}{\includegraphics{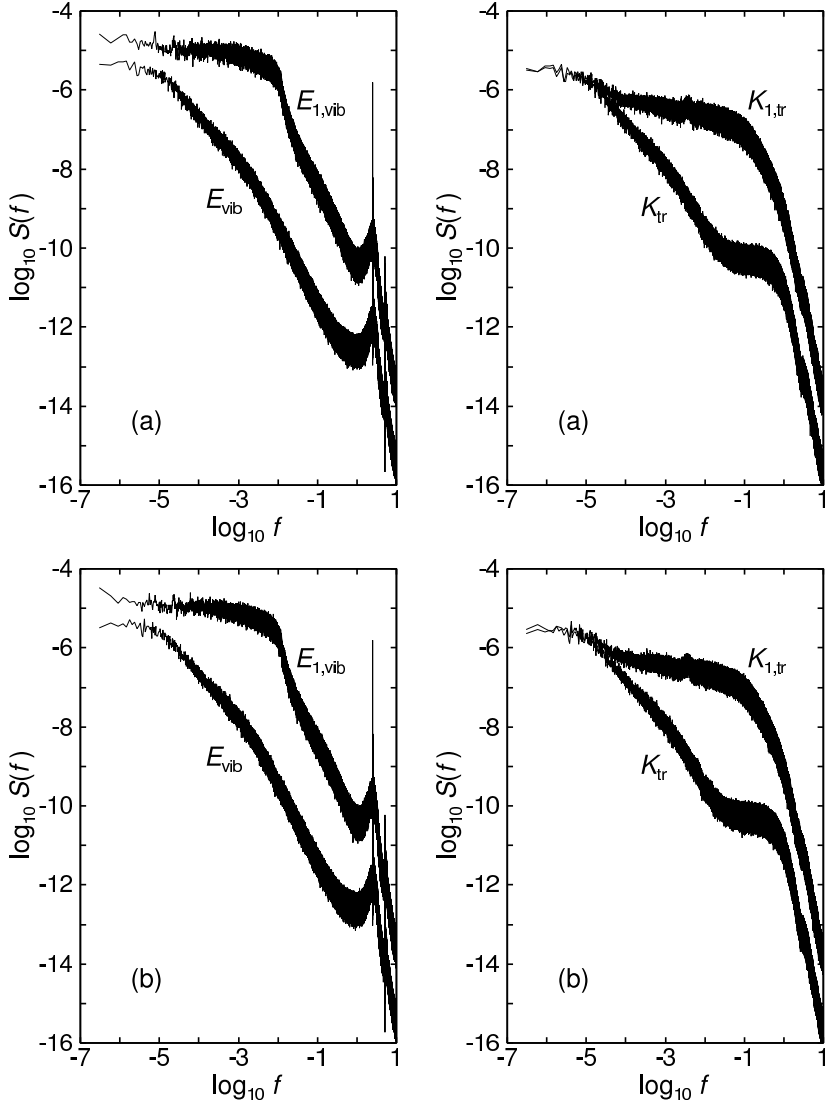}}
\caption{\label{fft-E1E2-128}
(a) Log-log plot of PSD for 
the vibrational energy: $E_{\rm vib}$ and for 
translational energy: $K_{\rm tr}$. 
The spectra for energies of a single molecule are denoted by 
$E_{1,{\rm vib}}$ and $K_{1,{\rm tr}}$.
$N=128, \, \omega=16, E_{\rm tot}=1.0$. 
The time interval of FFT time series data is 
$\Delta t =0.05$, 
the number of the FFT data is 
$N_{\rm dat}=2^{26}$,
i.e. the observation time is $t_{\rm obs}=3.4 \times 10^{6}$. 
Each PSD was obtained from the average over 20 independent runs 
with nearly equilibrium initial states. 
(b) The same as in (a) but with the initial states which are the final 
states of 20 runs in (a).
}
\end{figure}

The power spectrum density (PSD) $S(f)$ of energy fluctuations 
is presented in Fig. \ref{fft-E1E2-128}(a) for vibrational motion 
and for translational one, where
$\omega=16$, $N=128$, $E_{\rm tot}=1.0$ and 
the observation time $t_{\rm obs}=3.4 \times 10^{6}$.
The sharp peak at 
$f_{\rm vib} \equiv \omega/2\pi \approx 2.5$ 
in the spectrum of $E_{\rm vib}$
corresponds to the smallest time scale of the present system: 
$t_{\rm vib}=1/f_{\rm vib} \approx 0.39$.
(The second peak is its harmonics at $f=2f_{\rm vib}$.)

The next small time scale (apart from the time scale of the inter-molecular 
potential: $t_{\rm unit}=r_0 \sqrt{M/\phi_0}=1$) is the mean free time 
(collision time): 
$t_{\rm col}\equiv L/\overline v \approx 4.9$, where the molecular 
distance is $L=4$ and the average velocity of the molecule is 
$\overline v\approx 0.82$ because the average 
translational energy is 
$M{\overline v}^2/2=K_{\rm tr}\approx 1/3$. 
Corresponding to this time scale, a mild hump around 
$f_{\rm col} \equiv 1/t_{\rm col} \approx 0.21$ can be seen in the PSD of 
$K_{\rm tr}$.
 
\begin{figure}
\scalebox{0.55}{\includegraphics{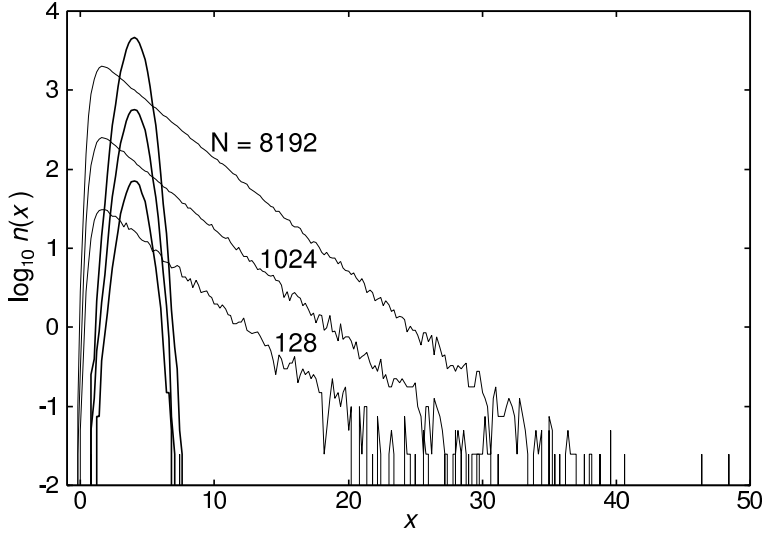}}
\caption{\label{disp-position}
Semilog plot of the distribution function $n(x)$ of molecular distance 
for $N=128, 1024$ and $N=8192$. $\omega=8, E_{\rm tot}=1.0$.  
$n(x)$ is normalized as $\int n(x) \, dx =N$. 
The initial distribution ($t=0$) and 
the distribution at $t=200$ are compared. At $t=0$, the distribution 
is gaussian with the center on $x=4$. 
The distribution at $t=200$ is almost final one and is well fitted 
to $\sim e^{-x/x_0}$ with $x_0 \sim 3$ for $x \gg x_0$. 
Each line was obtained from the average over 200 independent runs 
with nearly equilibrium initial states.
}
\end{figure}

We have employed the initial states in which 
the positions $q_i$ ($i=1, 2, \cdots N$) of  
molecules are distributed with Gaussian distribution 
with a small width around the equilibrium position as in (\ref{init-q-i}).
The distribution function $n(x)$ of the molecular distance 
$x=p_{i+1} -q_i$ is expected to tend to exponential: 
$n(x) \sim e^{-x/x_0}$ for large $x$ 
after a relatively short time which is long enough 
in comparison to $t_{\rm col}$. 
This is because the molecular interaction is short range repulsive and each 
molecule occupies its position on a line independently. 
That is confirmed by 
semilog plot of $n(x)$ for $N=128, 1024$ and $N=8182$ 
in Figure \ref{disp-position}. 
The initial gaussian distribution with the center $x=L$ tends to 
exponential with the average value $x_0 \approx 3$ as 
can be explained 
in terms of the average distance $L=4$ and the scale of 
the repulsive potential $r_0=1$ which yields $x_0 \sim L - r_0=3$. 

The spectra of $E_{\rm vib}$ and $K_{\rm tr}$ are flat at 
frequencies lower than a certain frequency at the ``shoulder'' which we denote 
$f_0$. In Fig. \ref{fft-E1E2-128}(a), the shoulder frequency is  
$f_0 \approx 10^{-5}$. 
To observe the saturated spectrum (the spectrum which is white 
at low frequencies) one needs to observe 
the system for a time interval at least 
$t_0 \equiv 1/f_0 \approx 10^5$. 
In other words, $t_0$ is a typical time scale after which the system 
loses the memory of the initial state. 
It should be pointed out that 
this long time scale characterizes 
the system in thermal equilibrium and 
does not depend on 
the specified initial state if chosen as an equilibrium one. 
To confirm it, 
we performed the simulation starting from the final state 
at $t=t_{\rm obs}$ of each run in (a) of Fig. \ref{fft-E1E2-128} 
and continued the simulation for another time interval $t_{\rm obs}$. 
The power spectra thus obtained 
were shown in (b) of Fig. \ref{fft-E1E2-128}, which is very similar to 
those in (a). 

We have defined $\overline {\tau}_{\rm cor}$ 
as a characteristic time scale of the system, which is 
$\overline {\tau}_{\rm cor} \approx 1.1 \times  10^{3}$ in the case of $N=128$, $\omega=16$ 
(see Fig. \ref{tau-omega-128}). 
If the correlation function is pure exponential (Debye-type relaxation) as 
$G(\tau)\sim e^{-2\pi f_{\rm c} \tau}$, the corresponding PSD 
is Lorentzian given by $S(f) \sim 1/[1+(f/f_{\rm c})^2]$ from 
the Wiener-Khinchin theorem and $S(f)$ is flat at  
low frequencies 
$f \lesssim f_{\rm c}$, i.e. $f_{\rm c}$ is the shoulder frequency.
If our $G(\tau)$ were exponential, this frequency should relate to 
the correlation time $\tau_{\rm cor}$ defined by (\ref{def-tau}) as 
$f_{\rm c}=$$\ln 2/2\pi \tau_{\rm col}$
$\sim \ln 2/2\pi \overline {\tau}_{\rm col}$
$\approx 1.0 \times 10^{-4}$. 
As we have seen in the above, 
the shoulder frequency 
$f_0$ is much smaller (at least more than one digit) than $f_{\rm c}$, 
which means 
that decay of the correlation function is slower than exponential, i.e. 
the PSD function deviates from Lorentzian. (See below.)

\begin{figure}
\scalebox{0.55}{\includegraphics{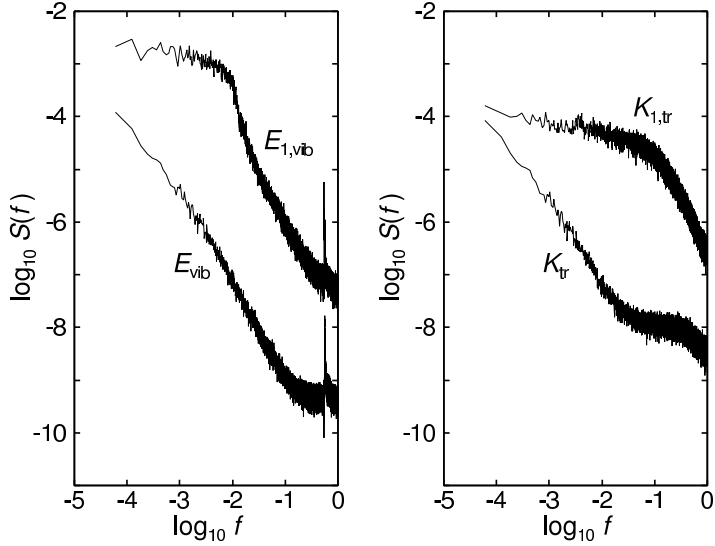}}
\caption{\label{fft-E1E2-128A}
The same as Fig. \ref{fft-E1E2-128} but with 
a shorter observation time: $t_{\rm obs}=1.6 \times 10^{4}$.
$\Delta t =0.5$ and $N_{\rm dat}=2^{15}$. 
The spectra of $E_{\rm vib}$ and $K_{\rm vib}$ is not yet 
saturated and look power-low like,  
while the spectra for the single molecule 
($E_{1,{\rm vib}}$ and $K_{1,{\rm vib}}$) are nearly flat in a low frequency 
range. 
}
\end{figure}
The long time scale appears clearly when one observes 
the energy involved with all molecules 
but not the energy of individual ones. 
The spectra of the vibrational and translational energies of molecule 1 
($E_{1,{\rm vib}}\equiv \pi_1^2/2M+M\omega^2 \xi_1^2/2$ 
and $K_{1,{\rm tr}}\equiv p_1^2/2M$) are 
also presented in Fig. \ref{fft-E1E2-128}. 
The shoulder frequencies in these spectra are 
much higher (order of three or four digits) than in 
the spectra of $E_{\rm vib}$ and $K_{\rm vib}$,  
implying that each molecule loses its initial memory 
much faster than the total system. 
Similar results for the energy power spectra of  
the total system and the individual molecules 
have been obtained from MD simulations of 
liquid water \cite{shudo05}. 

The long time scale is often related to the power law of PSD as
$\sim 1/f^{\alpha}$ with 
$0 < \alpha < 2$ over a wide range of frequency. 
It should be noted, however, that the spectrum sometimes appears to be 
$1/f^{\alpha}$-like 
when the observation time $t_{\rm obs}$ is not long enough. 
Power spectra obtained from 
the same simulations as in Fig \ref{fft-E1E2-128} but with a shorter 
observation time: $t_{\rm obs}\approx 1.6 \times 10^{4}$ and a longer 
sampling time $\Delta t=0.5$ are presented 
in Fig. \ref{fft-E1E2-128A} \cite{footnote4}.  
For energies of the single molecule, 
the power spectra are saturated,  
i.e. white at low frequencies, 
while the spectra of $E_{\rm vib}$ 
and $K_{\rm tr}$ look like 
$\sim 1/f^{\alpha}$ with $\alpha$ being $1.4 \sim 1.5$ 
over a low frequency range of about two digits and no shoulders are 
observed 
because the lowest frequency $1/t_{\rm obs}$ 
is higher than $f_0$.  
As mentioned above, these spectra deviate from the Lorentzian form
in which 
$\alpha =2$ should be expected for $f \gtrsim f_0$. 
The value of $\alpha$ smaller than two might be attributed to 
the heterogeneity, i.e. the system is 
more complex because of the internal degrees of freedom. 
In fact, 
MD simulations of much more complex systems \cite{shudo05}
(three-dimensional liquid water, alcohol and argon cluster)   
yielded power spectra whose exponent $\alpha$ of the power-law decaying 
part
is smaller than the present value.

\subsection{Relaxation from nonequilibrium initial states}
\label{subsec:from-noneq}

\begin{figure}
\scalebox{0.55}{\includegraphics{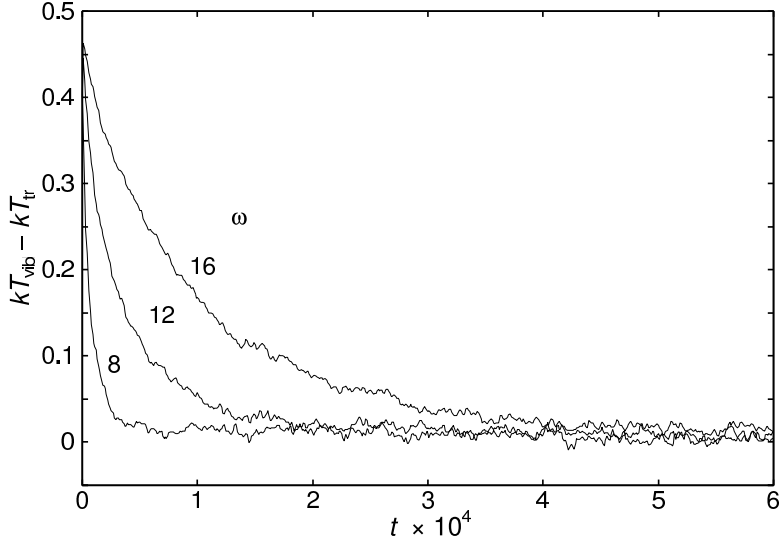}}
\caption{\label{temp-diff}
Temperature difference between the vibrational and 
translational motion. $N=128$. 
Temporal variation of 
$k \Delta T \equiv kT_{\rm vib}-kT_{\rm tr} \equiv E_{\rm vib}-2K_{\rm tr}$ 
is plotted for $\omega =8, \; 12, \; 16.$ 
Each line was obtained from the average over 200 independent runs with 
different initial states satisfying the condition 
$E_{{\rm tot},0}=1$, $K_{{\rm vib},0}=V_{{\rm vib},0}= 0.4$, 
$K_{{\rm tr},0}=0.2$, $V_{{\rm int},0}\approx 0$ so that the 
initial temperature is $kT_{{\rm vib},0}=0.8$, $kT_{{\rm tr},0}=0.4$. 
The plotted point at the time $t$ represents the average over 
the time interval $[t-100, t+100]$. The sampling interval of plotting is 
$\Delta t=100$. 
}
\end{figure}

\begin{figure}
\scalebox{0.55}{\includegraphics{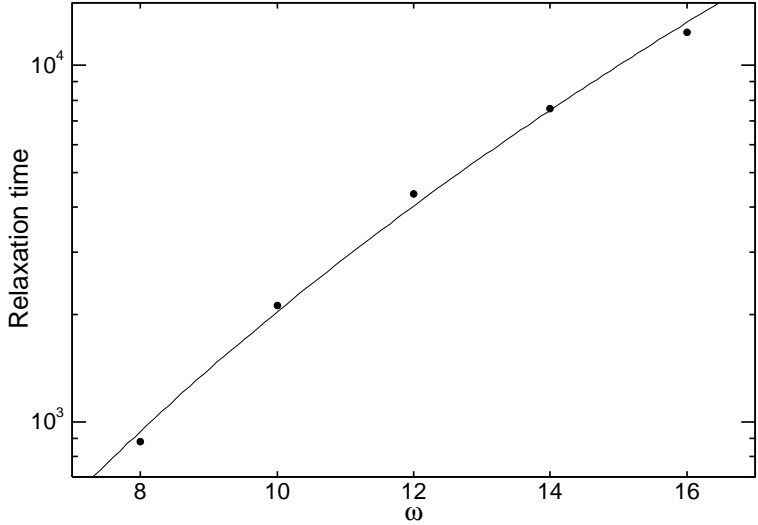}}
\caption{\label{relax-omega-dep}
Semilog plot of 
$\omega$-dependence of the relaxation time $\tau_1$ defined in 
Eq. (\ref{relaxation}).
$N=128, E_{\rm tot}=1.0$. The same initial condition as in 
Fig. \ref{temp-diff}.
The guide line is $A_1 \exp\,[B_1\omega^{0.4}\,]$ with 
$A_1=0.204$ and $B_1=3.601$. 
}
\end{figure}
In deriving the time scale in Eq. (\ref{tau-stretch-exp}) from 
the Landau-Teller 
approximation, a nonequilibrium initial state is assumed 
and the relaxation time toward equilibrium is estimated. 
Here we perform simulations based on this idea and initially 
prepare the system with different temperatures for subsystems of 
the translational and vibrational motion. 
We set the energies of subsystems 
as $K_{{\rm tr},0}=0.2$ and 
$E_{{\rm vib},0}\equiv K_{{\rm vib},0}+V_{{\rm vib},0}=0.8$ corresponding to 
the initial temperatures 
$kT_{{\rm tr},0}=2K_{{\rm tr},0}=0.4$ and 
$kT_{{\rm vib},0}=E_{{\rm vib},0}=0.8$. 
In Fig. \ref{temp-diff}, the temporal variation of the quantity 
$E_{\rm vib}-2K_{\rm tr}=k(T_{\rm vib}-T_{\rm tr})$ 
is presented for various values of $\omega$, which shows how the 
temperature difference of two subsystems tends to 
zero and the thermal equilibrium is realized. 

If the temperature difference $\Delta T\equiv T_{\rm vib}-T_{\rm tr}$ 
decays as
\begin{equation}
\frac{d}{dt}\Delta T (t) =-\frac{\Delta T (t) }{\tau_1},
\label{relaxation}
\end{equation}
the relaxation time $\tau_1$ derived from the Landau-Teller approximation 
depends on $\omega$ and $T_{\rm tr}$, 
and its main $\omega$-dependence is obtained as 
$\tau_1 = C\exp[B\omega^{\alpha}]$, where $C$ still depends 
on $\omega$ weakly and on $T_{\rm tr}$ \cite{benettin99}. 
If $T_{\rm tr}$-dependence of $C$ could be ignored, 
the lines in Fig. \ref{temp-diff} should be fitted to 
$k \Delta T(t) =k \Delta T(0) e^{-t/\tau_1}$ with  
$k\Delta T(0)=0.4$. 
In fact, these lines exhibit some oscillations as a function of the time $t$, 
even though each line has been obtained from the average over 
the time interval $[100-t, t+100]$ and from the average over 200 
independent runs.  
Especially the lines (for larger values of $\omega$) starting from 0.4 
do not decrease in the beginning but increase steeply  
(up to $\approx 0.47$ for $\omega =16$) \cite{footnote5}. 
Hence, the fitting to the 
exponentially decreasing function is very rough. We, however, estimated 
the relaxation time $\tau_1$ using the fitting function 
$0.4 \, e^{-t/\tau_1}$. The $\omega$-dependence of $\tau_1$ is given in 
Fig. \ref{relax-omega-dep} and the guide line is obtained from 
$\tau_1 = A_1 \exp \left[B_1 \omega^{0.4} \right]$ with $A_1=0.240$ and 
$B_1=3.601$. 
In comparison to the correlation time (see Fig. \ref{tau-omega-128}), 
data points of $\tau_1$ appreciably deviate from 
the fitting line. The main reason might be the temperature dependence 
of $\tau_1$ in Eq. (\ref{relaxation}) which has been ignored. 
The correlation time is estimated in thermal equilibrium, 
while the relaxation time here is defined in the nonequilibrium state, and 
the smaller the temperature difference, the better 
the approximation \cite{footnote6}. 

Next we compare the relaxation time $\tau_1$ 
(the time scale of energy exchange between the vibrational and 
the translational motion) and another time scale: the time scale of energy 
relaxation $within$ the vibrational degrees of freedom. 
we carried out a simulation starting from a nonequilibrium vibrational 
state, i.e. 
the initial state is  
such that the vibrational motion of 
subsystems I (composed of the first $N/2$ molecules) and 
II (composed of the rest $N/2$ molecules) are in different 
thermal equilibrium, 
while the translational motion is in another thermal equilibrium state 
as a whole.
The initial values of vibrational energies (per molecule) 
$E_{\rm I,vib},\; E_{\rm II,vib}$ 
of subsystems I and II were chosen as $E_{{\rm I,vib},0}=2E_{{\rm vib},0}$ and 
$E_{{\rm II,vib},0}=0$, respectively. 

\begin{figure}
\scalebox{0.55}{\includegraphics{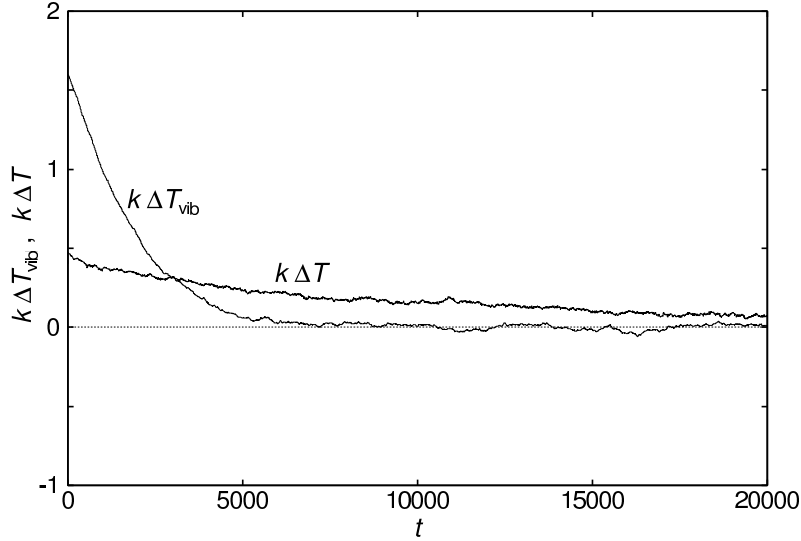}}
\caption{\label{temp-diff-1}
Temporal variation of the quantities 
$k \Delta T \equiv E_{\rm vib}-2K_{\rm tr}$ and 
$k \Delta T_{\rm vib} \equiv E_{\rm I,vib}-E_{\rm II,vib}$, 
where the suffixes I and II indicate the subsystems 
composed of the first $N/2$ and the rest $N/2$ molecules, respectively. 
Initial condition: 
$E_{{\rm tot},0}=1$, $E_{{\rm vib},0}=0.8$, 
$K_{{\rm tr},0}=0.2$, $V_{{\rm int},0}\approx 0$.
The vibrational energy is distributed only to the subsystem I
so that 
$K_{{\rm I, vib},0}=V_{{\rm I, vib},0}=0.8$,
and
$K_{{\rm II, vib},0}=V_{{\rm II, vib},0}=0$. Each line was 
obtained from the average over 20 independent runs satisfying the same 
initial condition. 
$N=256$, $\omega=16$. The sampling interval of plotting is 
$\Delta t=5$. 
}
\end{figure}

Fig. \ref{temp-diff-1} shows the case for 
$N=256$, $\omega=16$ and for the 
initial condition $E_{{\rm vib},0}=0.8$ and $K_{{\rm tr},0}=0.2$, where 
plotted points were obtained by the average over 20 different runs.
The temperature difference defined by 
$k \Delta T \equiv E_{\rm vib}-2K_{\rm tr}$, starting from 
$k \Delta T_0=0.4$, decays monotonically, 
while the temperature difference  
$k \Delta T_{\rm vib} \equiv E_{\rm I,vib}-E_{\rm II,vib}$ exhibits a   
damping oscillation. 
For a fixed value of $\omega$, the first zero-crossing time of 
$k \Delta T_{\rm vib}$ increases linearly with $N$, and the amplitude of 
the damping oscillation, i.e. the most negative value of 
$k \Delta T_{\rm vib}$ 
turned out to decrease with increasing $N$. On the other hand, 
for a fixed value of $N$, the first zero-crossing time of 
$k \Delta T_{\rm vib}$ increases weakly with increasing $\omega$ but 
$\tau_1$, i.e. the time scale of 
$k \Delta T$ increases much strongly ($\sim \exp^{B\omega^{0.4}}$). 
Thus, for sufficiently large values of both $N$ and $\omega$, 
the time scale of 
energy exchange between the translational 
and vibrational motion is much larger 
than that of energy relaxation within the vibrational system, 
which can actually be seen in 
Fig. \ref{temp-diff-1}.

\subsection{Effect of deviation from the complete resonance}
As mentioned already, the exponent $\alpha$ in Eq. (\ref{tau_B}) generally 
depends on the number $N$ of degrees of freedom, and the long time scale 
might not be realized for large $N$ except when all $\omega_i$'s are 
equal to each other.
Also in the Landau-Teller approximation, the binary collisions are 
such that the vibrational frequency of each molecule is 
identical. 
Hence it is interesting to investigate whether the condition of the complete 
resonance is essential for the long time scale in the system with
many degrees of freedom.
\begin{figure}
\scalebox{0.55}{\includegraphics{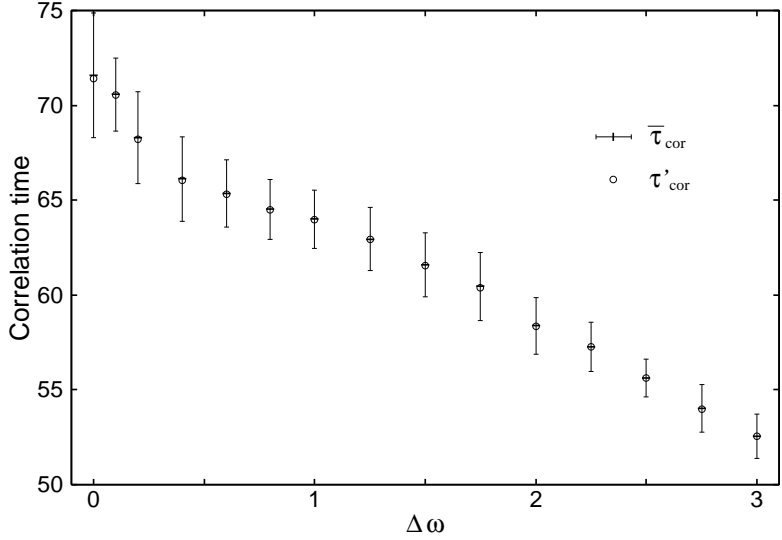}}
\caption{\label{del_omega-128}
$\Delta \omega$-dependence of the averaged 
correlation time $\overline{\tau}_{\rm cor}$. $N=128$, $E_{\rm tot}=1.0$.
The vibrational frequencies of molecules 
$(\omega_1, \cdots,  \omega_N)$ are generated from 
homogeneous random numbers with the width $\Delta \omega$ and 
a fixed average value $\omega=8$. 
$\overline{\tau}_{\rm cor}$ is 
obtained from the average over 40 independent runs with 
nearly equilibrium initial states. 
Error bars indicate $\overline{\tau}_{\rm cor} \pm \sigma$, 
where $\sigma$ is the standard deviation. 
The points of $\Delta \omega=0$ correspond to those of $N=128$ in 
Fig. \ref{tau-omega-128} but obtained from 40 runs.
}
\end{figure}

We have chosen homogeneous 
random numbers $(\omega_1, \cdots , \omega_N)$
in the interval $[\omega-\Delta \omega/2, \omega +\Delta \omega/2]$, whose 
average is $\omega$. 
Fig. \ref{del_omega-128} shows $\Delta \omega$-dependence 
of the correlation time for $N=128$, $E_{\rm tot}=1$, $\omega=8$. 
The correlation time decreased with increasing $\Delta \omega$ monotonically, 
but its $\Delta \omega$-dependence is rather complicated, 
which we have not yet investigated. 

\begin{figure}
\scalebox{0.55}{\includegraphics{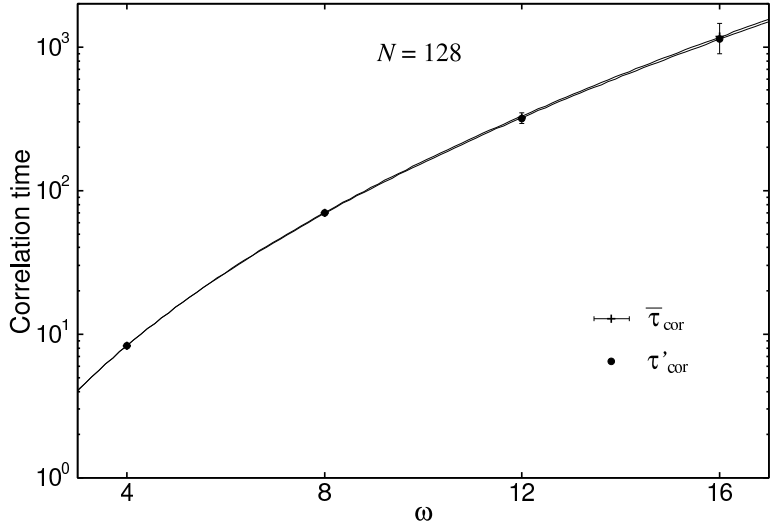}}
\caption{\label{deltaomega01}
Semilog plot of 
$\omega$-dependence of $\overline{\tau}_{\rm cor}$ for a fixed value of 
$\Delta \omega =0.1$. $N=128, E_{\rm tot}=1.0$. 
The average is taken over 40 independent runs with nearly equilibrium 
initial states and error bars express $\pm \sigma$. 
The guide line corresponds to 
$\overline{\tau}_{\rm cor} = A \exp[B \times \omega^{0.4}]$ 
with $A=0.0106, \; B=3.83$. 
}
\end{figure}

By assuming a fixed small value for $\Delta \omega$ as $\Delta \omega=0.1$, 
we obtained the $\omega$-dependence of the 
correlation time as in Fig. \ref{deltaomega01}.
The present result shows that 
the condition of complete resonance: $\omega_1= \cdots = \omega_N$ 
is not essential for the long time scale 
because the correlation time increases 
with $\omega$ in the same way as in $\Delta \omega=0$ 
for a finite (but small) value of $\Delta \omega$. 

\section{Conclusions and Discussions}

In conclusion, 
we estimated the correlation time $\tau_{\rm cor}$ of the vibrational 
energy which is defined in thermal equilibrium. 
The present simulations yielded the time scale which indeed remains finite 
in the large $N$ limit and grows as $\sim \exp{B \omega^{0.4}}$ 
in good agreement with that 
obtained from the Landau-Teller approximation. 
The condition of the complete resonance assumed both in the mathematical 
theorem based on the perturbation method and in the Landau-Teller 
approximation is not essential for the long time scale because a slight 
deviation from the complete resonance did not change the $\omega$-dependence 
of the time scale. 

Concerning the system size dependence, 
we found that the correlation time $\tau_{\rm cor}(N)$ 
decreased in the beginning but tends to a finite value with 
increasing $N$ for a fixed value of the molecular density. 
If the molecular density $\rho \equiv N/L_{\rm tot}=1/L$ 
($L_{\rm tot}$ is the length of the system) is small enough so that 
three-molecule collisions can be neglected, then one can 
expect $\tau_{\rm cor}(N)$ being constant (for a fixed value of $\rho$). 
In other words, the system size dependence of the time scale is regarded 
as arising from multi-molecular collisions. 
Then a natural question arises: 
Does the long time scale in the thermodynamic limit remains 
for high density case? 
The present paper does not answer this question.

The present study was done for a one-dimensional gas. 
To what extent does the one-dimensional scenario work in 
higher dimensional gases? 
On this issue, we note the case of FPU-type systems (as 
a model of lattice vibrations) \cite{footnote1}: 
The long time scale remains in a one-dimensional chain but disappeared 
in a two-dimensional lattice, 
which suggests the difficulty in predicting a correct answer 
for the gas model too. 
No systematic study of the size dependence has been reported for 
higher-dimensional systems including solids and liquid.
Nevertheless, our feeling is that, as suggested in \cite{shudo05}, 
if the system is heterogeneous 
(i.e. composed of subsystems with well separated time scales), 
the basic idea of Boltzmann and Jeans 
might be still usable and so the slow relaxation might be realizable in the 
thermodynamic limit for the high-dimensional gases as well
and even for more general systems covering liquid and solid.

The power spectra of the energy fluctuations obtained in the 
present simulations are $1/f^{\alpha}$-like 
with $\alpha$ around  $1.4 \sim 1.5$ in a range of low frequencies. 
In ``real'' $1/f$ fluctuations (or noise) observed in, say, 
the electrical resistance 
of condensed matter systems, 
$\alpha$ is nearly constant and very close to one over a wide range of 
frequencies. 
Typical upper limits of the power-law 
decaying range are from 100 to ${10^4}$Hz, 
and a typical lower limit is $10^{-2}$Hz. 
(In some case the lower limit 
has been reported to be extended down to $10^{-7}$Hz without observing 
the shoulder frequency.) \cite{dutta81}
Until now, such beautiful $1/f$ fluctuations have never been reproduced 
from any Hamiltonian systems with many degrees of freedom.
In case of one-dimensional FPU $\beta$-model, 
it was found \cite{fuchi96}, for large $N$, 
that 
the shoulder frequency 
$f_0$ of the lowest mode energy spectrum 
decreases with increasing $N$, 
while the spectrum is close to Lorentzian: $\alpha \approx 2$ in a  
frequency range $f \gtrsim f_0$. 
In other words, the energy fluctuations of the FPU $\beta$-model
cannot be said as $1/f$-type even if 
the time scale $\sim 1/f_0$ might be long. 
In contrast, the spectra for diatomic gas obtained here deviate 
from Lorentzian and have an exponent $\alpha$ smaller than two.  

Brillouin scattering experiments on quartz \cite{musha90} showed that phonon 
number
fluctuations (energy fluctuations of each normal mode) 
exhibit $1/f^{\alpha}$ spectra
over a frequency range from $10^{-4}$ to $10^{-1}$ Hz. 
In similar experiments on liquid water \cite{musha92}, 
$1/f^{\alpha}$ spectra were also observed, 
in which the frequency range is even wider: 
$10^{-4}$ to $10^{1}$ Hz.
In both cases, $\alpha$ is close to one and no shoulder appears 
during the observation time which is of orders of hours ($\sim 10^4$ s). 
As stated above, one-dimensional FPU models 
are unsatisfactory to explain such $1/f$ noise. 
On the other hand, modified FPU models having internal degrees of freedom
seem to be a better candidate. 
Actually, 
the analysis of a modified one-dimensional FPU model \cite{galgani92} is 
suggestive, in which the long time scale was proved to remain 
in the thermodynamic limit. 
At this stage, we think much of investigation into the power spectra of 
such systems 
to know whether the power-law decaying range could be wide enough 
in the thermodynamic limit and, more importantly, 
whether $\alpha$ could approach one
owing to the internal degrees of freedom.

{\bf Acknowledgments}

We are grateful to T. Kimura who collaborated with us at 
the early stage of this work.


\begin{thebibliography}{00}

\bibitem{boltzmann1895}
L. Boltzmann, Nature \textbf{51} (1895) 413.

\bibitem{jeans1903}
J. H. Jeans, 
Philos. Mag. \textbf{6} (1903) 279.

\bibitem{benettin87}
G. Benettin, L. Galgani and A. Giorgilli, 
Phys. Lett. A \textbf{120} (1987) 23.

\bibitem{benettin87(I)}
G. Benettin, L. Galgani and A. Giorgilli,
Commun. Math. Phys. \textbf{113} (1987) 87.
 
\bibitem{benettin89(II)}
G. Benettin, L. Galgani and A. Giorgilli,
Commun. Math. Phys. \textbf{121} (1989) 557.

\bibitem{galgani92}
L. Galgani, A. Giorgilli, A. Martinoli and S. Vanzini, 
Physica D \textbf{59} (1992) 334.

\bibitem{sasai92}
M. Sasai, I. Ohmine and R. Ramaswamy, J. Chem. Phys. \textbf{96} (1992) 3045.

\bibitem{ishijima98}
A. Ishijima, H. Kojima, T. Funatsu, M. Tokunaga, H. Higuchi, H. Tanaka 
and T. Yanagida,
Cell \textbf{92} (1998) 161.

\bibitem{shudo05}
A. Shudo and S. Saito, Adv. Chem. Phys. B \textbf{130} (2005) 375.

\bibitem{nakagawa01}
N. Nakagawa and K. Kaneko, Phys. Rev. E \textbf{64} (2001) 055205(R).

\bibitem{nekhoroshev77}
N. N. Nekhoroshev, Usp. Mat. Nauk\ \textbf{32} (1977). 
[Russ. Math. Surv. \textbf{32} (1977) 1.]

\bibitem{FPU55}
E. Fermi, J. Pasta and S. Ulam, Los Alamos Report No. LA-1940 (1955), 
in {\it Collected Papers of Enrico Fermi}, vol. II (Univ. Chicago Press, 1965) 
p. 978.

\bibitem{benettin86}
G. Benettin, 
Proc. Int. School of Phys. ENRICO FERMI, Course XCVII, 
ed. G. Ciccotti and W. G. Hoover (1986) p. 15.

\bibitem{berchialla04}
L. Berchialla, A. Giorgilli and S. Paleari, Phys. Lett. A 
\textbf{321} (2004) 167.

\bibitem{benettin05}
G. Benettin, Chaos \textbf{15} (2005) 015108.


\bibitem{benettin94}
G. Benettin, Prog. Theor. Phys. Suppl. No. 116 (1994) 207.

\bibitem{benettin99}
G. Benettin, P. Hjorth, and P. Sempio, 
J. Stat. Phys. \textbf{94} (1999) 871.

\bibitem{landau36}
L. Landau and E. Teller, Phys. Z. Sowjet. \textbf{10} (1936) 34, in 
{\it Collected Papers of L. D. Landau}, ed. D. Ter Haar 
(Pergamon Press, 1965) p. 147. 

\bibitem{rapp60}
D. Rapp, J. Chem. Phys. \textbf{32} (1960) 735.

\bibitem{yoshida90}
H. Yoshida, Phys. Lett. A \textbf{150} (1990) 262.

\bibitem{press02}
W. H. Press, S. A. Teukolsky, W. T. Vetterling and B. P. Flannery, 
Numerical Recipes in C++, 2nd ed. pp. 205.
(Cambridge Univ. Press, Cambridge, UK. 2002)

\bibitem{dutta81}
P. Dutta and P. M. Horn, 
Rev. Mod. Phys. \textbf{53} (1981) 497.

\bibitem{fuchi96}
H. Kawamura, N. Fuchikami, D. Choi and S. Ishioka,
Jpn. J. Appl. Phys. \textbf{35} (1996) 2387.

\bibitem{musha90}
T. Musha, G. Borbely and M. Shoji, 
Phys. Rev. Lett. \textbf{64} (1990) 2394.

\bibitem{musha92}
T. Musha and G. Borbely, 
Jpn. J. Appl. Phys. \textbf{31} (1992) L370.


\bibitem{footnote1}
In fact, the numerical 
simulation of the FPU $\beta$-model showed 
the long time scale as $\sim \exp[(1/\varepsilon)^{0.25}]$ (here 
$\varepsilon$ is the total energy divided by $N$) 
in the limit $N \to \infty$ \cite{berchialla04}, 
while in the FPU $\alpha$-model \cite{galgani92} and also in a two-dimensional 
FPU lattice \cite{benettin05}, 
the non-ergodic behavior disappeared. 

\bibitem{footnote2}
The system here is the same as that employed in \cite{benettin87} except 
the boundary condition and maybe the details of the initial condition.

\bibitem{footnote3}
Even though random numbers are generated such that 
their probability distribution function $P(x)$ 
satisfies $\int x P(x) dx =0$, 
this does not mean that the average value of these numbers 
actually vanishes.

\bibitem{footnote4} 
The peaks at 
$f=f_{\rm vib} -1/\Delta t \approx 0.55$ in the spectra of $E_{\rm vib}$ and
$E_{1,{\rm vib}}$ 
are an artifact arising from the aliasing effect \cite{press02}.

\bibitem{footnote-tau-fit}
Note that the 
parameters $(A,B)$ and $(A',B')$ do not need to coincide 
because those parameters depend on the initial state of each run. 
Here the correlation times were obtained from 
20 random initial conditions and the average values
$\overline{\tau}_{\rm cor}$ and $\tau_{\rm cor}'$ are  
defined in different ways.



\bibitem{footnote5} 
It turned out that this increase corresponds to the situation that the initial 
energy transfer occurs mainly from $K_{\rm tr}$ to $V_{\rm int}$,
while $E_{\rm vib}$ stays almost constant.

\bibitem{footnote6} 
Remembering the non-exponential decay of $G(\tau)$, we like to mention 
another possibility: 
The time-dependence of the temperature difference $\Delta T$ 
might deviate from pure exponential.



\end{thebibliography}
\end{document}